\documentclass[twocolumn]{aastex701}
\usepackage{enumitem}

\newcommand{\rev}[1]{\textcolor{black}{#1}}

\begin{document}

\title{Accretion is All You Need: Black Hole Spin Alignment in Merger GW231123 \\ Indicates Accretion Pathway}

\author{Imre Bartos}
\affiliation{Department of Physics, University of Florida, PO Box 118440, Gainesville, FL 32611-8440, USA}
\email[show]{imrebartos@ufl.edu}  

\author{Zolt\'an Haiman} 
\affiliation{Institute of Science and Technology Austria (ISTA), Am Campus 1, Klosterneuburg, Austria}
\affiliation{Department of Astronomy, Columbia University, 550 West 120th Street, New York, NY, 10027, USA}
\affiliation{Department of Physics, Columbia University, 550 West 120th Street, New York, NY, 10027,
USA}
\email[show]{zoltan.haiman@ista.ac.at}

\begin{abstract}
GW231123 represents the most massive binary-black-hole merger detected to date, lying firmly within, or even above, the pair-instability mass gap. The component spins are both exceptionally high ($a_1 = 0.90^{+0.10}_{-0.19}$, $a_2 = 0.80^{+0.20}_{-0.51}$), which is difficult to explain with repeated mergers. Here we show that the black hole spin vectors are closely \emph{aligned with each other} while significantly tilted relative to the binary’s orbital angular momentum, pointing to a common accretion-driven origin. We examine astrophysical formation channels capable of producing near-equal, high-mass, and mutually aligned spins consistent with GW231123---particularly binaries embedded in AGN disks and Pop~III remnants, which grew via coherent misaligned gas accretion. We further argue that other high-mass, high-spin events, e.g., GW190521 may share a similar evolutionary pathway. These findings underscore the critical role of sustained, coherent accretion in shaping the most extreme black hole binaries.
\end{abstract}

\section{Introduction}\label{sec:intro}

The detection of the gravitational--wave event GW231123\_135430 (hereafter GW231123; \citealt{2025arXiv250708219T}) by the LIGO \citep{2015CQGra..32g4001L}, Virgo \citep{VIRGO:2014yos}, and KAGRA \citep{KAGRA:2018plz} observatories revealed the most massive binary black‑hole (BH) coalescence observed to date, with component masses\footnote{Throughout this {\it Letter}, we quote the median and symmetric 90\% credible intervals reported by \cite{Abbott2025GW231123}.}
%
$m_1 = 137^{+22}_{-17}\,M_\odot$ and $m_2 = 103^{+20}_{-52}\,M_\odot$. Equally remarkable are the large, dimensionless spin magnitudes
$\chi_1 = 0.90^{+0.10}_{-0.19}$ and
$\chi_2 = 0.80^{+0.20}_{-0.51}$, both of which were found to be significantly tilted relative to the binary’s orbital angular momentum \citep[]{Abbott2025GW231123}.  

The high mass of GW231123 is difficult to reconcile with stellar‐evolution calculations that predict the suppression of BH formation within the mass range
$\sim 60$--$130\,M_\odot$ due to pair‑instability pulsation supernovae (PPSN) and pair‑instability supernovae (PISN)  (\citealt{Heger2002,WoosleyHeger2021}; albeit see \citealt{2020ApJ...905L..15B,Farmer2019,Marchant2020} that found that the gap may narrow, due to, e.g., extreme $^{12}$C$(\alpha,\gamma)^{16}$O and other reaction rates). This motivates consideration of non-stellar channels for producing such heavy BHs. 

There are two main pathways to grow BH masses above those of stellar remnants: hierarchical mergers or sustained accretion. Hierarchical mergers, in which a merger remnant undergoes subsequent coalescences \citep{GerosaBerti2017,2017ApJ...840L..24F,2019PhRvL.123r1101Y,Tagawa+2021-hierarchical}, have been proposed to explain some of the heaviest observed BH mergers, in particular GW190521 \citep{Abbott2020GW190521,2021ApJ...920L..42G}. Such mergers are typically expected in dense star clusters where BHs can come together by chance, or in the disks of active galactic nuclei (AGN) where interaction with the disk can bring BHs into a smaller volume, acting as a BH assembly line.  Since recoil velocities after a merger are typically 100s of ${\rm km~s^{-1}}$, hierarchical assembly is most likely in environments with escape velocities above this value -- in practice, this requires mergers to occur near a nuclear supermassive BH (SMBH).

Several recent studies have proposed hierarchical origins for GW231123.  \citet{Stegmann+2025-GW231123} explore binary capture in a nuclear star cluster where spin alignment leads to high remnant spin and low recoil, facilitating repeated mergers, while \citet{Li+2025-GW231123,Kiroglu+2025} model a succession of mergers involving $\sim10$ stellar-mass BHs.

Sustained accretion has been previously proposed in two scenarios: (1) in AGN disks where BHs are within the dense disk environment (\citealt{Levin2007,McKernan+2012,Bartos2017,Stone2017,2018ApJ...859L..25Y,Secunda2019,2020ApJ...901L..34Y}); and (2) in the case of formation from metal‑poor or metal‑free ({\sc Pop\,III}) stars, where the newly formed BHs continue to grow by accreting gas in the dense cores of protogalactic halos~\citep{RoupasKazanas2019,2020ApJ...903L..21S}.

The hierarchical merger and accretion channels can in principle both explain the extreme masses of GW231123, but they make distinct predictions for spin magnitudes and orientations. In the former case, second‑generation BHs produced by previous mergers have characteristic spins $\chi\sim0.7$. Reproducing $\chi\gtrsim0.8$ spins requires fine-tuned mass ratios and ancestral BH spins \citep{PhysRevD.81.084023,2025arXiv250708219T}.

Accretion, on the other hand, can spin a BH up rapidly. While radiative capture allows spins up to $\chi\lesssim0.998$ \citep{Thorne1974}, strong magnetic fields and disc winds impose an empirical ceiling of $\chi\sim0.9$ for sustained accretion \citep{Reynolds2021}. Indeed, X‑ray reflection studies of AGNs frequently measure SMBH spins in this range (\citealt{Reynolds2021,2024arXiv240612096W}). The two spins in GW231123 thus sit at this putative limit.

Beyond the magnitude of BH spins, their orientation also carries information of the binary's origin and evolution. Importantly, it may also help differentiate between hierarchical and accretion formation processes. In hierarchical formation scenarios, the progenitor BHs form independently prior to binary assembly, so their spin vectors are expected to be uncorrelated. By contrast, co-evolution through accretion within a binary system can lead to BH spins that are aligned with each other.  In the most naive picture, in which the circumbinary disk and the binary orbit are coplanar, the BH spins are also aligned with their orbital angular momentum~\citep{Bogdanovic2007}, but as we argue below, a significant spin-orbit misalignment can naturally be expected.

For all previous LIGO-Virgo-KAGRA BH mergers, the modestly high spin magnitudes and signal-to-noise ratios made the relative angle $\theta_{12}$ between the two BH spins effectively unmeasurable \citep{Vitale2014}.  
The high spins and signal-to-noise ratio (SNR\,$>20$)  of GW231123, however, make it an interesting target for measuring $\theta_{12}$ for the first time.%

In this {\it Letter}, we examine the {\em relative} spin geometry of GW231123. We quantify the observable correlation between the two BH spins and interpret it---together with other properties of the merger---in the context of accretion-driven evolution.

\section{Relative Spin Alignment}
\label{sec:spin_alignment}

\begin{figure}[h]
\centering
\includegraphics[width=\columnwidth]
{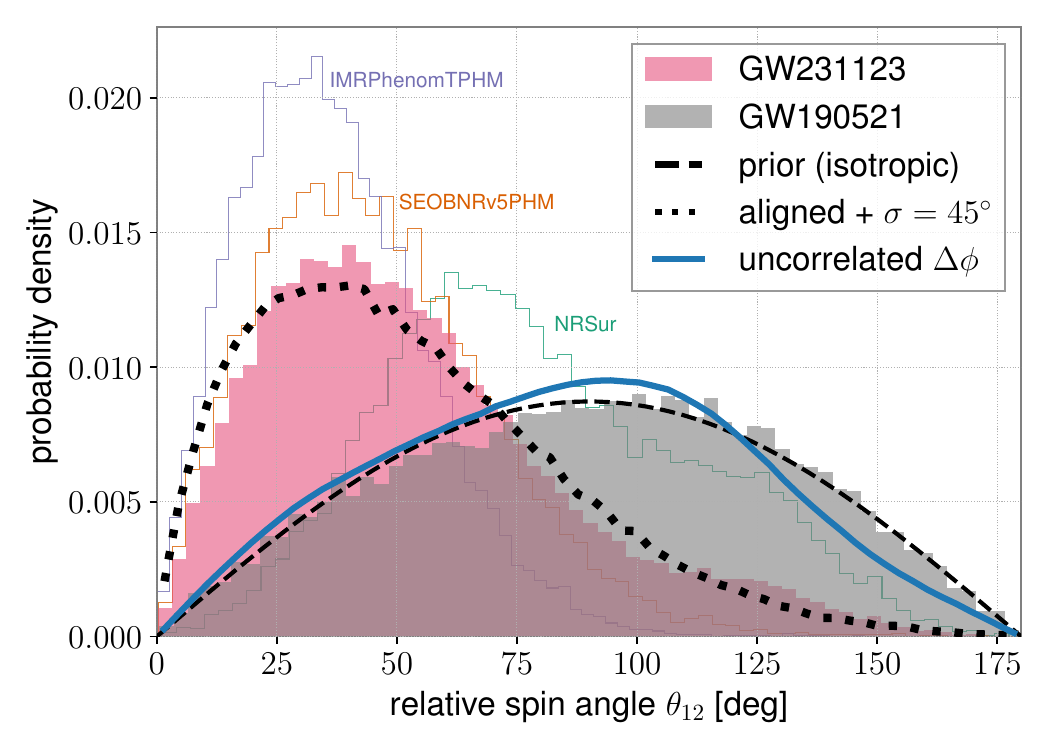}
\caption{{\bf Posterior distribution of  the relative spin angle $\theta_{12}$ for GW231123} (red histogram) and GW190521 (gray histogram), compared to theoretical expectations. The dashed black line shows the isotropic prior, $p(\theta_{12}) = \frac{1}{2} \sin \theta_{12}$. The dotted black line shows the distribution expected for perfectly aligned spins observed with Gaussian tilt uncertainty of \rev{$\sigma = 45^\circ$}. The blue curve shows the distribution obtained by randomly sampling the azimuthal angle $\Delta\phi$ while preserving the measured tilt angles from GW231123. \rev{Also shown are the three model distributions used to obtain the GW231123 posterior.}}
\label{fig:histograms}
\end{figure}

To investigate the relative spin orientation of the BHs in GW231123, we computed the angle $\theta_{12}$ (not to be confused with the tilt angles $\theta_1$ and $\theta_2$) between the two spin vectors using posterior samples from gravitational-wave parameter estimation. These samples were obtained from the public data release associated with the event, available via Zenodo\footnote{\url{https://doi.org/10.5281/zenodo.16004263}}, and correspond to the model-averaged (``combined’’) posterior provided by the LIGO–Virgo–KAGRA Collaboration, incorporating multiple waveform models including \texttt{IMRPhenomXPHM}, \texttt{SEOBNRv4PHM}, and \texttt{NRSur7dq4}.

From the posterior sample set we extracted the three spin orientation parameters for each component: the spin magnitudes $\chi_1$ and $\chi_2$, the tilt angles $\theta_1$ and $\theta_2$ (defined as the angle between each spin and the orbital angular momentum), and the azimuthal angle difference $\Delta\phi = \phi_2 - \phi_1$. 
The relative spin angle $\theta_{12}$ was computed from these quantities for each posterior sample using
\begin{equation}
\cos \theta_{12} = \cos \theta_1 \cos \theta_2 + \sin \theta_1 \sin \theta_2 \cos \Delta\phi.
\end{equation}
We evaluated this expression across the entire posterior sample set to construct a posterior distribution for $\theta_{12}$.

The resulting distribution is shown in Fig. \ref{fig:histograms}. We also show the isotropic prior distribution, along with the posterior distribution we obtain for GW190521 (similarly using publicly available data; \citealt{Abbott2020GW190521}). We see that while the distribution of GW190521 does not meaningfully deviate from the prior, GW231123 strongly tilts towards low angles, suggesting some level of alignment.

To gauge how tightly the spins must be aligned, we compare the data with a synthetic population in which both black-hole spins are drawn from a Gaussian cone of width $45^{\circ}$ ($1\sigma$) around the same, but otherwise arbitrary, reference axis; each pair is then assigned a relative azimuthal angle $\Delta\phi$ chosen uniformly in $[0,2\pi)$.  We see that this distribution is reasonably close to that of GW231123, indicating an alignment within $\lesssim45^\circ$.

A further pathway to apparent alignment occurs when both spins are dominated by components parallel to the orbital axis: these parallel parts alone force $\theta_{12}$ to be small, and because aligned spin components are well constrained, such axis-aligned configurations are readily measurable even if the in-plane spin components remain poorly determined.

For GW231123 this is not expected to be the case since both spins are mostly perpendicular to the orbital axis with small measured aligned spin components ($\chi_{\rm eff}$; \citealt{2025arXiv250708219T}). To quantify this, we carried out a simulation in which we kept the posterior tilt angles ($\theta_1$,$\theta_2$) for GW231123 but randomized $ \Delta\phi$ drawing it from a uniform distribution. Fig. \ref{fig:histograms} shows the resulting distribution (in blue). We see that this is consistent with isotropic spin orientations, showing that our result for $\theta_{12}$ is not due to alignment with the orbital axis.

\section{Properties from sustained accretion}

Accretion can lead to distinct BH properties that are worth examining both for their consistency with GW231123 and possibly with other, earlier LIGO-Virgo-KAGRA detections, as well as for expectations for future detections that could more clearly distinguish this channel.
\newline

\noindent{\bf Mass:} Long-term accretion can substantially increase the mass of
BHs. Under Eddington-limited accretion, the BH mass grows exponentially with
time as
\rev{
\begin{equation}
M(t) = M_0 \exp\!\left( f_{\rm Edd}\, \frac{t}{t_{\rm Sal}} \right),
\end{equation}
where $M_0$ is the initial mass, $f_{\rm Edd}\equiv \dot M/\dot M_{\rm Edd}$ is
the Eddington ratio,} and
\begin{equation}
t_{\rm Sal} = \frac{\epsilon\,c\,\sigma_T}{4\pi G m_p}
\approx 45\,{\rm Myr}\left(\frac{\epsilon}{0.1}\right)
\end{equation}
is the Salpeter time for radiative efficiency $\epsilon$. For example, a BH of mass $30\,{\rm M}_\odot$ that resides in an AGN disk and accretes at the Eddington rate for the lifetime of the disk (10--100\,Myr) will reach a mass of $37$--$280\,{\rm M}_\odot$. This can easily accommodate the masses of GW231123 and suggests that even higher masses are possible.

In fact, in the inner regions of the accretion disks of bright quasars (i.e. with near-Eddington luminosities for the SMBH), the growth can be more rapid. The rate at which gas is gravitationally captured by stellar-mass BHs can be many orders of magnitude higher than their Eddington rates in these regions (see, e.g., Fig. 2 in \citealt{Stone2017}).  Under such super- or hyper-Eddington fueling conditions, most of the incoming mass is expected to be lost to radiatively-driven outflows~\citep[e.g.][]{YuanNarayan2014}. Nevertheless, radiation-hydrodynamical simulations have found that at sufficiently high fueling rates, the BH growth remains super-Eddington, 
i.e. $\dot{M}_{\rm bh}=17.1 \dot{M}_{\rm Edd}(\dot{m_0}/{300})^{1/2}$, where 
$\dot{M}_{\rm Edd}\equiv L_{\rm Edd}/c^2$ is the accretion rate corresponding to the Eddington luminosity and $\dot{m_0}$ is the outer fueling rate of the BH from large radii (see Eq.~25 in \citealt{Hu+2022-super-Eddington})in units of $\dot{M}_{\rm Edd}$ (see Eq.~25 in \citealt{Hu+2022-super-Eddington}).  As an example, a stellar-mass BH in the inner $\lesssim 0.01$pc of the accretion disk of a bright quasar with a $\sim 3\times 10^6~{\rm M_\odot}$ SMBH can grow at $\gtrsim 100$ times its Eddington rate.
\newline

\noindent{\bf Mass ratio:} Accretion in binary BH systems tends to preferentially increase the mass of the lighter companion (e.g., \citealt{Farris+2014,2015MNRAS.452.3085Y,Duffell+2020, Siwek+2023}). This occurs because gas in the circumbinary disk is preferentially funneled toward the smaller BH, which is displaced farther from the center of mass and is orbiting closer to the "wall" of a circumbinary cavity.
This effect tends to equalize the two masses, therefore, even if the two BHs are formed separately and form a binary by chance, the observed mass ratio in accretion-driven binaries can be close to unity. This is consistent with GW231123, which has a mass ratio of $0.75^{+0.22}_{-0.39}$. \rev{Importantly, a mass ratio approaching unity suppresses the differential precession of the two spins, allowing mutual spin alignment to remain stable against gravitational-wave driven evolution even at large separations.}
\newline

\noindent{\bf Spin magnitude:} Sustained disk accretion can efficiently amplify BH spin magnitudes, driving them toward near-maximal values. In the case of thin-disk accretion with aligned angular momentum, the spin evolution is governed by the change in BH angular momentum as mass is added from the innermost stable circular orbit (ISCO). The spin-up follows the relation
\begin{equation}
\frac{d\chi}{dm} = \frac{1}{M} \left( \frac{l_{\rm ISCO}(\chi)}{e_{\rm ISCO}(\chi)} - 2\chi \right),
\end{equation}
where $l_{\rm ISCO}$ and $e_{\rm ISCO}$ are the specific angular momentum and energy of the infalling gas at the ISCO, respectively \citep{Bardeen1970,Thorne1974}. Integrating this equation shows that to spin up a non-rotating BH to $\chi \approx 0.9$, approximately $1.7$ times its initial mass must be accreted.

\begin{figure}[h]
\centering
\includegraphics[width=\columnwidth]
{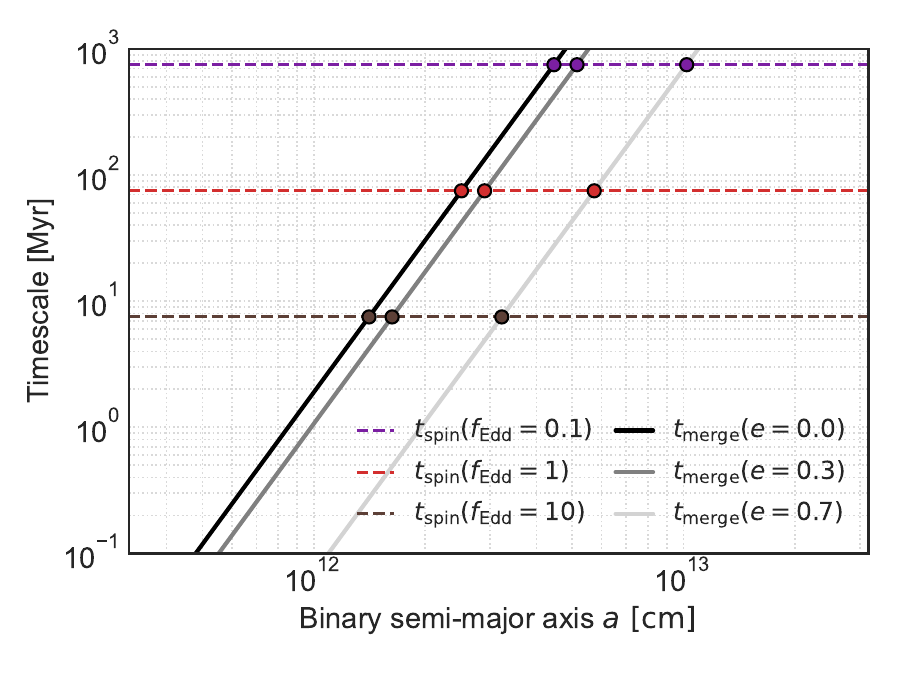}
\vspace{-0.6cm}
\caption{\rev{Gravitational-wave merger time $t_{\rm merge}$ for a
GW231123-like binary as a function of its semi-major axis $a$ for
eccentricities $e = 0, 0.3, 0.7$ (black to light gray curves).  Horizontal
dashed lines show the spin-up times $t_{\rm spin}$ required to reach
$\chi \simeq 0.9$ for Eddington ratios $f_{\rm Edd}=0.1,1,10$.  The
markers indicate the separations at which $t_{\rm merge} = t_{\rm spin}$,
defining a minimum semi-major axis above which gas accretion has
sufficient time to spin the black holes up before gravitational-wave
dominated inspiral.}}
\label{fig:spinup}
\end{figure}

\rev{Using the Salpeter growth time defined above, the time required to accrete the
$\Delta M \simeq 1.7 M_0$ needed to reach $\chi\simeq 0.9$ is $t_{\rm spin} \simeq \frac{t_{\rm Sal}}{f_{\rm Edd}}\approx 45\,{\rm Myr}\, f_{\rm Edd}^{-1}$. For a binary system, this spin–up time must be shorter than the remaining lifetime of the binary.  A useful diagnostic is to compare $t_{\rm spin}$ with the gravitational–wave merger time $t_{\rm merge}(a,e)$ \citep{1964PhRv..136.1224P}. Figure~\ref{fig:spinup} shows $t_{\rm merge}$ for a GW231123-like binary as a function of its semi–major axis for several eccentricities, together with the spin–up times for $f_{\rm Edd}=0.1$, 1, and 10.  Their intersections define a minimum separation $a_{\min}$ above which accretion has sufficient time to reach $\chi\simeq0.9$ before gravitational--wave--dominated inspiral takes over.}  

\rev{In one of the most comprehensive semi-analytic models to date \cite{Tagawa2020}, binaries form in AGN disks at distances of $r\gtrsim 0.01$pc from the SMBH, with initial separation comparable to their mutual Hill radius, or $(m/M)^{1/3}r \gtrsim 10^{14}$cm.  We note that for binaries at sufficiently large separations, torques from circumbinary gas may speed up the inspiral.  While the timescale for this gas-assisted inspiral $t_{\rm gas}$ is highly uncertain, two-dimensional hydrodynamical simulations find that it is comparable to the mass accretion time-scale $t_{\rm acc}=M/(dM/dt)$ \citep{2020ApJ...900...43T,2020ApJ...901...25D}.  As a result, in case gas is important for the inspiral, the spin-up, accretion, and inspiral timescales are all expected to be comparable, allowing for both growth and spin-up to occur before merger.}

In the case of Pop III stellar origin there is essentially no such time constraint and, therefore, sub-Eddington accretion is still consistent with spin magnitudes $\chi>0.8$ (on the other hand, growth of Pop III stellar remnant BHs may be limited by these BHs moving out of dense regions; e.g. \citealt{Regan+2020}).
\newline

\noindent{\bf Spin orientation:} Accretion through individual \emph{minidisks} can efficiently funnel gas onto the BHs, leading to spin-up. If the gas flow is coplanar with the binary orbital plane, the angular momentum of the inflowing gas tends to align the BH spins with the binary angular momentum, leading to spin directions approximately perpendicular to the orbital plane \citep{Bardeen1970,Bogdanovic2007}.

However, this picture can be more complex in realistic astrophysical environments. Hydrodynamic simulations have shown that when gas accretes from a misaligned circumbinary disk, the minidisk planes themselves can become misaligned with the binary orbit, depending on the disk thickness and the angular momentum transport mechanisms \citep{2012PhRvD..85h4015L}. \citet{Nixon2013} find that for thin disks with $H/R \sim 0.01$, sustained misaligned accretion can maintain tilted minidisks, and hence tilted spin orientations. In contrast, \citet{Moody2019} simulate thicker disks ($H/R \sim 0.1$) and observe realignment of the minidisk and circumbinary disk planes with the binary orbit, suggesting that alignment is more efficient in more viscous environments.

Planet formation analogs also illustrate the possibility of polar accretion, even when the accretion occurs from a co-planar disk. For example, \citet{Szulagyi2022} show that in unequal-mass systems like Jupiter and its circumplanetary environment, gas accretes onto the planet from polar directions—above and below the plane of the protoplanetary disk. This suggests that if a single planet is replaced by a binary whose orbital plane is aligned with that of an AGN disk, accretion onto the binary can likewise occur from the polar directions.

Recent theoretical work further complicates the picture by considering spin–orbit coupling effects. \citet{Ennoggi2025} argue that misaligned spins can lead to precession of the minidisks and non-planar accretion flows due to gravitomagnetic torques. They suggest that close to coalescence, in-plane spin components may cause ``bobbing'' of the orbital plane—a general relativistic effect discussed in \citet{Koppitz2007}—potentially introducing shocks and variability in the accretion flow structure.

More generally, when a binary accretes gas coherently but from a misaligned direction, it is much easier to realign the BH spins than the binary’s orbital plane. For accretion onto a single BH, the final spin direction $\hat{J}_{\rm spin}$ tends to align with the total angular momentum vector $\vec{J}_{\rm tot} = \vec{J}_{\rm gas} + \vec{J}_{\rm spin}$, where $\vec{J}_{\rm gas}$ is the cumulative angular momentum of the inflowing gas \citep{King+2005}. Analogously, for a binary system, the orbital angular momentum $\vec{J}_{\rm bin}$ will tend to align with $\vec{J}_{\rm tot} = \vec{J}_{\rm gas} + \vec{J}_{\rm bin}$. However, because the BH spins carry far less angular momentum than the binary orbit -- typically by a factor of $(a/R_{\rm g})^{1/2} \gg 1$, where $a$ is the binary separation and $R_{\rm g}$ is its gravitational radius -- the spins can align with the gas flow much earlier during the accretion process. This implies that during misaligned accretion, there exists a phase where the BH spins are aligned with each other and with the angular momentum of the inflow, but remain misaligned with the binary orbital plane.

Taken together, the above considerations suggest that while prolonged accretion often drives spin alignment, factors such as misaligned inflow, disk geometry, and relativistic couplings can sustain or even amplify spin misalignments-producing a broader diversity of spin–orbit configurations than the simplest alignment scenarios would predict.

\section{What Could Differentiate Between AGN and Pop~III Origins?}

The mutually aligned but orbit-misaligned spins in GW231123 can be explained by both AGN and Pop~III accretion scenarios, but key differences in spin–orbit dynamics and mass distributions offer potential discriminants between these channels.
\newline


\noindent{\bf Orbital Tilt and Misalignment:} AGN-assisted binaries are subject to close encounters that can tilt the orbital plane independently of the BH spins \citep{Tagawa2020}. This can enhance spin–orbit misalignment while preserving mutual spin alignment, consistent with the configuration observed in GW231123.

In contrast, Pop~III binaries evolve in relative isolation, making the binary plane aligned with the accretion disk. If the spin tilt in such cases is limited (e.g. to $\lesssim 40^\circ$ in \citealt{2012PhRvD..85h4015L}), then a higher spin tilt favors an AGN origin.
\newline


\noindent{\bf Mass--Spin Relationship as a Diagnostic:} In the Pop~III channel the binary evolves in relative isolation, without strong third-body perturbations or a dense stellar environment. Any post-formation accretion proceeds from a single, quasi-fixed reservoir (e.g., residual envelope or a long-lived circumbinary flow), so the angular-momentum direction of the inflow remains approximately constant. Under such \emph{coherent} accretion, each BH spins up monotonically as its mass grows, yielding a tight, monotonic mass--spin relation. 
Assuming an initial mass $M_0 \!=\! 50\,M_\odot$, a BH that reaches $\chi \!\sim\! 0.9$ typically attains a mass of $\sim\!120\,M_\odot$. Hence, if GW231123-like systems originate from Pop~III remnants, all BHs with $M \gtrsim 120\,M_\odot$ are expected to have near-maximal spins ($\chi\!\gtrsim\!0.8$--$0.9$).

Binaries embedded in AGN disks migrate through a torque-rich, time-variable medium, so the inflow direction changes (turbulence, spiral waves, encounters at migration traps), intermittently tilting the minidisks. Successive episodes then add angular momentum with differing orientations, weakening cumulative spin-up. Consequently, even very massive BHs in AGN disks can retain sub-maximal spins, yielding a broad high-mass mass--spin distribution—unlike the tight Pop~III trend.
\newline


\noindent{\bf Expected Mass Distributions:} The expected mass spectrum of progenitor BHs differs significantly between the two channels. Pop~III stellar evolution predicts a relatively flat or even top-heavy initial mass function for remnants due to weak stellar winds and direct-collapse pathways \citep{Susa2014,Hirano2015}, though the precise slope remains model dependent. In contrast, AGN-assisted mergers are expected to originate from a dynamically segregated population within nuclear star clusters. This implies a steeper initial mass function---possibly a power law of the form $dN/dM \propto M^{-1.5}$ \citep{Tagawa2020}.

The implication for observability is that, under Pop~III origins, the LIGO-Virgo-KAGRA network should preferentially detect the heaviest mergers, since the detection volume increases with chirp mass. In contrast, AGN channels can produce both low- and high-mass events, with the lower-mass population still detectable in sizable numbers. These low-mass binaries may also show the characteristic signature of spin--spin alignment without requiring maximal spins, offering a possible observational signature of AGN-assisted accretion.
\newline


\noindent{\bf Orbital eccentricity:} Orbital eccentricity can be enhanced in AGN-assisted mergers by binary-single interactions in the galactic center \citep{2022Natur.603..237S,Gayathri2022}. Since these interactions can also tilt the orbital axis while they leave spin directions virtually unaffected, in the AGN scenario orbit-spin tilt may be correlated with detectable orbital eccentricity.
\newline

\noindent\rev{{\bf Expected rates for GW231123-like events:} Although AGN-assisted mergers are expected to contribute only a subset of the total BH merger rate (e.g., 5--30\% in \citealt{2021ApJ...920L..42G}), this level is consistent with an accretion-driven pathway operating across all such systems.  The most distinct observational signatures of accretion, namely high masses and high spins, are expected to appear most clearly in the upper end of the AGN-grown population where prolonged disk-driven growth is most effective.  The highest-mass observed mergers, including GW231123 and GW190521, display these characteristic features.
In contrast, the rate and properties of Pop III binaries remain highly uncertain.}
\newline

\noindent\rev{\noindent{\bf Tidal disruptions:} A third pathway that can produce high spins through sustained accretion involves young and gas-rich stellar clusters.  In these environments, stellar-mass black
holes can undergo repeated interactions with massive stars, including collisions
and tidal disruptions, which provide prolonged and coherent mass transfer. Several recent studies have shown that such episodes can efficiently spin up BHs and, under certain conditions, lead to partial spin alignment
\citep{2025ApJ...983L...9K,2025ApJ...979..237K,2023MNRAS.525.5752R,2023MNRAS.524.6358K,2019ApJ...877...56L}. Although the predicted rates remain uncertain, these cluster-driven accretion episodes represent an additional channel capable of producing high-spin
mergers.}
\newline 

\noindent\rev{\noindent{\bf Spin--orbit coupling and the stability of mutual alignment.}
Post-Newtonian evolution admits a resonant configuration where the two spins remain locked in a mutual alignment ($\Delta\phi \simeq 0$) while precessing jointly about the total angular momentum \citep{2004PhRvD..70l4020S,2020PhRvD.101l4037M}. However, the stability of this resonance at large separations depends critically on the binary properties. For unequal-mass systems, the differential spin--orbit coupling (which scales as $\propto r^{-2.5}$) dominates over the spin--spin coupling (the "locking" force, scaling as $\propto r^{-3}$) at large separations ($r \gtrsim 50\,M$). Consequently, if a binary with the median mass ratio of GW231123 ($q \approx 0.74$) were to decouple from the gas disk at typical decoupling radii ($r \sim 100\,M$), differential precession would destroy the alignment, randomizing the relative spin angle before the spins could lock. 
The observed alignment in GW231123 therefore implies one of two scenarios specific to the accretion channel. First, the system may have a mass ratio close to unity ($q \gtrsim 0.9$), which is well within the posterior support. As $q \to 1$, the symmetry of the system suppresses differential precession, rendering the resonant alignment stable even at large separations. Alternatively, if the masses are unequal, the gas torques must persist deep into the inspiral phase. Hydrodynamic simulations suggest that streams of gas can penetrate the circumbinary cavity \citep{2014ApJ...783..134F}, potentially maintaining the alignment torques down to small separations where the vacuum spin--spin coupling becomes strong enough to take over and preserve the lock.}
\section{Conclusion and Outlook}

We examined the possibility that accretion in a gas-rich environment led to the formation of the exceptionally massive, high-spin BBH merger GW231123. By analyzing the relative spin orientation of the two BHs, we found evidence for mutual spin alignment---a configuration naturally explained by coherent accretion, but more difficult to reconcile with repeated mergers. We discussed how sustained accretion in AGN disks or from Pop~III gas reservoirs can produce both the high spins and masses observed, and highlighted potential ways to distinguish between these two channels in future observations. Our key takeaways are as follows.
\begin{itemize}[left=0pt,itemsep=2pt,parsep=0pt]
    \item GW231123 is the first gravitational-wave event for which azimuthal spin alignment between the two BHs can be measured, enabled by its high spins, tilted spin orientations, and high signal-to-noise ratio.  We find significant alignment.
    
    \item Sustained accretion naturally explains nearly all unique features of GW231123: component masses beyond the pair-instability limit; near-equal mass ratio; high spin magnitudes and mutual spin alignment.
    
    \item High spins misaligned from the orbital axis may also be common in gas-rich environments, especially if binaries accrete from a circumbinary disk tilted with respect to the orbital angular momentum. Similar features in events like GW190521 suggest they may also belong to this accretion-driven channel and should motivate studies to better understand the  angular momentum distribution of gas accreted by binaries in three-dimensional simulations.

    \item Three diagnostics could distinguish between AGN-assisted and Pop~III accretion: (i) detection of high-spin mergers at lower masses (expected only in AGNs); (ii) a consistent $\sim0.9$ spin for all high-mass BHs (expected only in the Pop~III case); and (iii) in the AGN scenario, binary eccentricity may be correlated with spin tilt with respect to the orbital axis.
    
    \item \rev{Achieving both high mass and $\chi \sim 0.9$ spin in AGN disks is naturally explained by prolonged accretion, potentially including super-Eddington phases in the inner regions of bright disks.}
    


\end{itemize}

\begin{acknowledgments}
The authors thank Davide Gerosa  and Matthew Mould for valuable suggestions. We are grateful for support by the National Science Foundation under grant No. PHY-2309024 (IB) and by NASA under grants 80NSSC22K0822 and 80NSSC24K0440 (ZH). We used OpenAI’s ChatGPT \citep{openai2022chatgpt} during the preparation of this manuscript. This material is based upon work supported by NSF's LIGO Laboratory which is a major facility fully funded by the National Science Foundation. 
\end{acknowledgments}

\bibliography{Refs}{}
\bibliographystyle{aasjournal}
\end{document}